# Simultaneous Metal-Insulator and Antiferromagnetic Transitions in Orthorhombic Perovskite Iridate $Sr_{0.94}Ir_{0.78}O_{2.68}$ Single Crystals

H. Zheng[1], J. Terzic[1], Feng Ye[2,1], X.G. Wan[3], D. Wang[3], Jinchen Wang[1,2,5], Xiaoping Wang[6], P. Schlottmann[4], S. J. Yuan[1] and G. Cao[1*]

[1]Center for Advanced Materials and Department of Physics and Astronomy

University of Kentucky, Lexington, KY 40506, USA

[2]Quantum Condensed Matter Division, Oak Ridge National Laboratory,

Oak Ridge, Tennessee 37831, USA

[3]Department of Physics, Nanjing University, Nanjing, China

[4]Department of Physics, Florida State University, Tallahassee, FL 32306, USA

[5]Department of Physics, Renmin University of China, Beijing, China

[6]Chemical and Engineering Materials Division, Oak Ridge National Laboratory,

Oak Ridge, Tennessee 37831, USA

The orthorhombic perovskite $SrIrO_3$ is a semimetal, an intriguing exception in iridates where the strong spin-orbit interaction coupled with electron correlations tends to impose a novel insulating state. We report results of our investigation of bulk single-crystal $Sr_{0.94}Ir_{0.78}O_{2.68}$ or Ir-deficient, orthorhombic perovskite $SrIrO_3$. It retains the same crystal structure as stoichiometric $SrIrO_3$ but exhibits a sharp, simultaneous antiferromagnetic (AFM) and metal-insulator (MI) transition at 185 K. Above it, the basal-plane resistivity features an extended regime of almost linear-temperature dependence up to 800 K but the strong electronic anisotropy renders an insulating behavior in the out-of-plane resistivity. The Hall resistivity undergoes an abrupt sign change and grows below 40 K, which along with the Sommerfeld constant of 20 mJ/mole K$^2$ suggests a multiband effect. All results including our first-principles calculations underscore a delicacy of the metallic state in SrIrO$_3$ that is in close proximity to an AFM insulating state. The contrasting ground states in isostructural Sr$_{0.94}$Ir$_{0.78}$O$_{2.68}$ and SrIrO$_3$ illustrate a critical role of even slight lattice distortions in rebalancing the ground state in the iridates. Finally, the observed simultaneous AFM and MI transitions reveal a direct correlation between the magnetic transition and formation of a charge gap in the iridate, which is conspicuously absent in Sr$_2$IrO$_4$.

*Introduction* It is now widely recognized that strong spin-orbit interaction (SOI) coupled with the electron-electron interaction (on-site Coulomb repulsion U) drives novel narrow-gap Mott insulating states in iridates **[1-7]**. The SOI is a relativistic effect proportional to $Z^2$ (Z is the atomic number), and is approximately 0.4 eV in the iridates (compared to ~ 20 meV in 3d materials), and splits the $t_{2g}$ bands into states with $J_{eff}$ = 1/2 and $J_{eff}$ = 3/2, the latter having lower energy **[1-7]**. Since $Ir^{4+}$ ($5d^5$) ions provide five 5d-electrons to bonding states, four of them fill the lower $J_{eff}$ = 3/2 bands, and one electron partially fills the $J_{eff}$ = 1/2 band where the Fermi level $E_F$ resides. The $J_{eff}$ = 1/2 band is so narrow that even a reduced U (~ 0.5 eV, due to the extended nature of 5d-electron orbitals) is sufficient to open a small gap Δ supporting the insulating state in the Ruddlesden-Popper series, $Sr_{n+1}Ir_nO_{3n+1}$ (n = 1 and 2; n defines the number of Ir-O layers in a unit cell). The splitting between the $J_{eff}$ = 1/2 and $J_{eff}$ = 3/2 bands narrows and the two bands progressively broaden and contribute to the density of states near the Fermi surface as the dimensionality (i.e., n) increases in $Sr_{n+1}Ir_nO_{3n+1}$. In particular, the 5d-bandwidth W of the $J_{eff}$ = 1/2 band increases from 0.48 eV for n = 1 to 0.56 eV for n = 2 and 1.01 eV for n = ∞ **[2, 4, 6]**. Since the effects of the SOI and U are largely local and remain essentially unchanged throughout the entire series, $Sr_{n+1}Ir_nO_{3n+1}$, the ground state evolves with decreasing Δ and increasing W, from a robust antiferromagnetic (AFM) insulating state for $Sr_2IrO_4$ (n = 1) to a paramagnetic semimetallic state for $SrIrO_3$ (n = ∞). That such a semimetallic state occurs in $SrIrO_3$ notwithstanding the presence of the strong SOI is of great interest both theoretically and experimentally **[2, 7-21]**. It is conceived that the strong SOI reduces the threshold of U for a metal-insulator (MI) transition **[8, 9]**. More recently, theoretical studies find that the stronger SOI requires a larger critical U instead for a MI transition to occur in SrIrO$_3$ because of a protected Dirac node for given U in the J$_{eff}$=1/2 bands near the Fermi level due to a combined effect of the lattice structure and strong SOI **[10, 11]**. In essence, small hole and electron pockets with a low density of states take place in SrIrO$_3$ and render a less effective U, which would otherwise drive a magnetic insulating state **[10, 11]**. Indeed, tuning the relative strength of the SOI and U effectively changes the ground state in the iridates, and the rare occurrence of a semimetallic state in SrIrO$_3$ provides an unique opportunity to closely examine the intricate interplay of the SOI, U and lattice degrees of freedom. The correlation between the AFM state and MI transition in the iridates has been among the most discussed topics in recent years **[7, 22, 23]**.

A good number of excellent studies of the orthorhombic perovskite SrIrO$_3$ have been conducted in recent years **[2, 13-21]**. However, bulk orthorhombic perovskite SrIrO$_3$ forms only at high pressures and high temperatures **[12, 13]**, and almost all studies of SrIrO$_3$ so far are limited to samples in thin-film or polycrystalline form **[2, 13-21]**. Little work on bulk single-crystal SrIrO$_3$ has been reported and critical information, such as anisotropies, magnetic properties and their correlation with the electronic state, etc., is still lacking. (Note that single crystals of the *hexagonal* SrIrO$_3$, which have been studied **[24]**, are not part of this work.) In this paper, we report results of our study on bulk single-crystals of the orthorhombic perovskite Sr$_{0.94}$Ir$_{0.78}$O$_{2.68}$ for an extended temperature range from 0.05 K to 800 K. Sr$_{0.94}$Ir$_{0.78}$O$_{2.68}$ retains the very same crystal structure as stoichiometric SrIrO$_3$ but exhibits sharp, concurrent AFM and MI transitions at 185 K with a charge gap of 0.027 eV, sharply contrasting with stoichiometric SrIrO$_3$ that is paramagnetic and semimetallic. The electrical resistivity is highly anisotropic and features an extended regime of approximately linear-T basal-plane resistivity between 185 K and 800 K. The Hall resistivity undergoes an abrupt sign change near 40 K (rather than at $T_N$=185 K), from holelike at high temperatures to electronlike at low temperatures, which along with a finite Sommerfeld constant of 20 mJ/mole K$^2$ implies in-gap states. Our band structure calculations confirm a small energy gap and finite magnetic moment in an AFM insulating state as a result of the Ir deficiencies in $Sr_{0.94}Ir_{0.78}O_{2.68}$. This work underscores the delicacy of the metallic state that is in close proximity to an AFM insulating state that is highly sensitive to even slight lattice defects in $SrIrO_3$. The concurrent AFM and MI transitions provide clear evidence for a direct correlation between the magnetic transition and charge gap in $SrIrO_3$, which is absent in $Sr_2IrO_4$.

*Experimental details* The single crystals studied were grown at ambient pressure from off-stoichiometric quantities of $SrCl_2$, $SrCO_3$ and $IrO_2$ using self-flux techniques. The formation of nonstoichiometric $SrIrO_3$ single crystals at ambient pressure is consistent with recent observations that the orthorhombic perovskite phase can be stabilized at ambient pressure via chemical doping **[13-16]**. The crystal structure of single-crystal $Sr_{0.94}Ir_{0.78}O_{2.68}$ was determined using a Rigaku X-ray diffractometer XtaLAB PRO equipped with PILATUS 200K hybrid pixel array detector at the Oak Ridge National Laboratory. More than 20 crystals were carefully examined. Full data sets were collected at 150K and the structures were refined using SHELX-97 and FullProf software **[25, 26]** (see **Table 1**). Chemical compositions of single-crystal $Sr_{0.94}Ir_{0.78}O_{2.68}$ were determined using energy dispersive X-ray analysis (EDX) (Hitachi/Oxford 3000) and x-ray diffraction discussed below. Magnetization, specific heat, Hall and electrical resistivity were measured using either a Quantum Design MPMS-7 SQUID Magnetometer and/or Physical Property Measurement System with 14-T field capability. The Hall resistivity as functions of temperature or magnetic field was determined by the difference between two sets of transverse resistivity data taken at a magnetic field with the same magnitude but opposite directions. AC resistance bridges (Linear Research 700 and Lakeshore 370) were used for transport measurements. The high-temperature resistivity was measured using a Displex closed cycle cryostat capable for a continuous temperature ramping from 9 K to 900 K.

The electronic band structure calculations were performed using the full potential linearized augmented plane wave method as implemented in WIEN2K package. Local spin density approximation (LSDA) for the exchange-correlation potential was used. We utilized an LSDA+SOI+U (U = 3 eV) scheme. A large cell used for the calculations is 4x unit cell. The self-consistent calculations were considered converged when the difference in the total energy of the crystal does not exceed 0.01 mRy.

**Table 1. Structural parameters for single-crystal perovskite $Sr_{0.94}Ir_{0.78}O_{2.68}$ at 150K**

| T = 150 K | The lattice parameters: $a$ = 5.541(6) Å, $b$ = 5.535(5) Å, $c$ = 7.833(8) Å, and $V$ = 240.3(4) Å$^3$; Space group: **Pbnm (No. 62)**. The agreement factor $R_1$ = 4.33% was achieved by using 392 unique reflections with I > 4σ and resolution of $d_{min}$ =0.65 Å. Anisotropic atomic displacement parameters were used for all elements involved. | | | | | |
|---|---|---|---|---|---|---|
|  | Site | x | y | z | Occupancy | Ueq(Å$^2$) |
| Sr1 | 4c | 0.0029(2) | 0.5094(2) | 0.25 | 0.94(3) | 0.0174(5) |
| Ir2 | 4c | 0.0029(2) | 0.5094(2) | 0.25 | 0.06(3) | 0.0174(5) |
| Ir1 | 4a | 0 | 0 | 0 | 0.72(3) | 0.0086(3) |
| O1 | 4c | -0.067 (2) | -0.004(1) | 0.25 | 1 | 0.022(3) |
| O2 | 8d | 0.232(1) | 0.268(2) | 0.035(1) | 0.84(4) | 0.022(2) |

*Crystal structure* The single-crystal $Sr_{0.94}Ir_{0.78}O_{2.68}$ adopts an orthorhombic perovskite structure with space group *Pbnm* (No. 62) having lattice parameters $a$ = 5.541(6) Å, $b$ =

5.535(5) Å, $c$ = 7.833(8) Å, and V = 240.3(4) Å$^3$, shown in **Fig. 1** and **Table. 1**. It has 3.08 Ir atoms, rather than 4, in a unit cell but its crystal structure is consistent with that of stoichiometric, orthorhombic perovskite SrIrO$_3$ reported in literature **[12-17]**. The Ir and O deficiencies inevitably lead to proportionally smaller lattice parameters of Sr$_{0.94}$Ir$_{0.78}$O$_{2.68}$ than those of SrIrO$_3$ with an approximate 2.5% decrease in volume (for the high-pressure bulk SrIrO$_3$ phase**,** $a$ = 5.60075 (5.58871) Å, $b$ = 5.57115 (5.57245) Å, $c$ = 7.89601 (7.88413) Å, and V = 246.376 (245.534) Å$^3$ at 300 K (3 K) **[17]**). The Ir-O-Ir bond angles for in-plane and out-of-plane are 161.1 (3)$^o$ and 162.3 (4)$^o$, respectively. A close comparison with the data of the stoichiometric SrIrO$_3$ in Ref. 17 indicates that Sr$_{0.94}$Ir$_{0.78}$O$_{2.68}$ has a larger in-plane rotation of IrO$_6$ octahedra, 9.54$^o$ (vs. 8.75$^o$) but a smaller out-of-plane tilt, 8.85$^o$ (vs. ~12$^o$). The in-plane rotation is likely to have a greater impact on physical properties than the out-of-plane tilt. Note that the IrO$_6$ octahedra in Sr$_2$IrO$_4$ and Sr$_3$Ir$_2$O$_7$ only rotate within the basal plane by about 12$^o$ without the out-of-plane tilt **[27-31]**.

One of the central features of Sr$_{0.94}$Ir$_{0.78}$O$_{2.6}$ is that a robust AFM transition occurring at T$_N$=185 K (**Fig.2a**) is accompanied by a sharp MI transition. Fitting the a-axis magnetic susceptibility χ$_a$(T) in **Fig. 2a** to a Curie-Weiss law for 200-340 K yields the Curie-Weiss temperature θ$_{CW}$ = +160 K and the effective moment μ$_{eff}$ = 0.19 μ$_B$/Ir. Like that of Sr$_2$IrO$_4$ and other related iridates **[4, 31, 32]**, θ$_{CW}$ show a characteristic positive sign, despite the AFM ground state. μ$_{eff}$ is finite but significantly smaller than that for other more insulating iridates (*e.g.*, 0.4 μ$_B$/Ir for Sr$_2$IrO$_4$) due to the more itinerant nature of the system discussed below. The isothermal magnetization M(H) varies almost linearly with the magnetic field H up to 7 T, as anticipated for an AFM ground state (see

**Fig.2b**). The magnetic anisotropy is visible but weak, consistent with the nature of the orthorhombic perovskite. For comparison, the evolution of the AFM order in the Ruddlesden-Popper series $Sr_{n+1}Ir_nO_{3n+1}$ with n=1, 2 and $Sr_{0.94}Ir_{0.78}O_{2.6}$ (n=∞) is illustrated in **Fig. 2c**. $T_N$ for $Sr_{0.94}Ir_{0.78}O_{2.6}$ is clearly lower than that for n=1 (240 K) and n=2 (285 K) but the low-temperature magnetic moment of $Sr_{0.94}Ir_{0.78}O_{2.6}$ seems considerably stronger than that of n=2 (right scale in **Fig.2c**).

The electrical resistivity $\rho(T)$ features a sharp MI transition at $T_N$=185 K and an almost linear-T dependence of the a-axis $\rho_a$ for an extended regime of 185 K-800 K. Along the c-axis, the apparent electronic anisotropy leads to a much less metallic behavior in which $\rho_c$, with much weaker temperature dependence above $T_N$, is more than one order of magnitude greater than $\rho_a$ despite the nature of the perovskite structure (see **Fig.3a**; note that $\rho_c$ is on a logarithmic scale on the right axis). The c-axis $\rho_c$ fits perfectly an activation law, $\rho_c \sim \exp(\Delta/T)$, for 125 K <T< 350 K; an evident slope change at $T_N$ signals a significant widening of the charge gap $\Delta$ from 0.008 eV above $T_N$ to 0.027 eV below $T_N$ (see **Inset** in **Fig.3a**). $\rho_c$ does not follow any power law below 125 K. The occurrence of the sharp MI transition at $T_N$ with the pronounced change of $\Delta$ clearly indicates a strong correlation between the AFM transition and charge gap in this iridate. This behavior sharply contrasts with that of $Sr_2IrO_4$ in which such a correlation seems more intricate, unconventional and still open to debate **[22, 23]**.

An extended regime of linear temperature resistivity is a classic signature of high-$T_C$ cuprates and the p-wave superconductor $Sr_2RuO_4$, Fe-based superconductors and many other correlated oxides **[33]**, in which spin fluctuations play an important role in the electron scattering. Elementary Bloch-Grüneisen theory predicts $\rho(T) \sim T$ for T >>

$\theta_D$, in the case of electron-phonon scattering where Debye temperature $\theta_D$ = 343 K for $Sr_{0.94}Ir_{0.78}O_{2.68}$. Resistivity saturation is anticipated when the mean-free path $l$ of the quasiparticles becomes shorter than the lattice parameter $a$ (Mott-Ioffe-Regel limit [34, 35]), or for $\rho \sim$ 100-150 $\mu\Omega$ cm (Mooij limit, [36, 37]). Here $\rho_a(T)$ for $Sr_{0.94}Ir_{0.78}O_{2.68}$ is well above the Mott-Ioffe-Regel or Mooij limits, and yet shows no sign of saturation up to 800 K. Furthermore, both $\rho_a(T)$ and $\rho_c(T)$ show an anomaly near T*=40 K that accompanies no corresponding magnetic anomaly. The origin of T* is unclear but closely related to the Hall resistivity discussed below.

For comparison and contrast, $\rho_a$ for $Sr_{n+1}Ir_nO_{3n+1}$ with n=1, 2 and $Sr_{0.94}Ir_{0.78}O_{2.6}$ (n=∞) is presented in **Fig. 3b**. $\rho_a$ decreases rapidly by as much as seven orders of magnitude at low temperatures as n increases from n=1 to n=∞, and the ground state evolves from the insulating state at n=1 to a much more metallic state at n=∞, particularly above $T_N$.

The Hall resistivity $\rho_H(T)$ taken at $\mu_oH$=7 T for $Sr_{0.94}Ir_{0.78}O_{2.68}$ exhibits an abrupt change in temperature dependence near T*, marking a drop in $\rho_H(T)$, thus a sign change from positive to negative with decreasing temperature, suggesting a multiband effect (**Fig. 3c**), which is qualitatively consistent with observations in thin-film $SrIrO_3$ [20, 21]. It appears that holes as charge carriers predominate above T* whereas electrons become overwhelming below T*. Moreover, $\rho_H(H)$ is negative with a strong dependence on magnetic field H at low temperatures and becomes positive with a much weaker field dependence at high temperatures (see **Inset** in **Fig.3c**). A two-carrier type model (p and n) needs to be considered to adequately determine the carrier densities but a rough estimate using the Drude theory yields the carrier densities to be of order of $10^{19}/cm^3$, suggesting a significantly enhanced itinerancy, compared to ~$10^{17}$/cm$^3$ for Sr$_2$IrO$_4$ **[38]**.

A few features in $\rho_H$ are also remarkable. In a magnetic system, the Hall resistivity consists of the ordinary and extraordinary Hall resistivity, empirically expressed by $\rho_H = R_oB + \mu_oR_eM$, where $R_o$ and $R_e$ are the ordinary and extraordinary Hall coefficient, respectively, B the applied magnetic field, $\mu_o$ the magnetic permeability of free space and M the magnetization. The ordinary Hall resistivity due to the Lorentz force on the conduction electrons tends to be temperature-independent whereas the extraordinary Hall resistivity usually requires the presence of localized moments and is associated with the exchange interaction between localized moments and the conduction electrons or the coupling of the orbital angular momentum L of the charge carriers to the spin angular momentum S of the scattering center or localized moments; therefore the extraordinary Hall resistivity is normally temperature-dependent. Here $\rho_H(T)$ is essentially temperature-independent above T* and then rapidly decreases with decreasing temperature. It is intriguing that the scattering of the electrons with local magnetic moments becomes pronounced only below T* rather than near the AFM transition $T_N$=185 K. It seems that the changes in $\rho_H$ track the magnetic fluctuations but are not necessarily coupled to a long-range magnetic transition in the iridate. This behavior is also seen in Sr$_2$IrO$_4$ **[39]**.

The specific heat C(T) was measured over an extended range, 0.05 K < T < 250 K. A slope change in C(T) occurs near $T_N$, confirming the magnetic phase transition (see **Fig. 4a**). C(T) for 0.05 K < T < 4 K approximately fits the common expression, $C(T) = \gamma T + \beta T^3$, where the first term arises from the electronic contribution to C(T) and the second term the phonon contribution; $\gamma$ is usually a measure of the density of states of the conduction states near the Fermi surface and effective mass, however, localized states or tunneling of atoms in a double well potential can also contribute to $\gamma$. $\beta$ is related to the Debye temperature $\theta_D$, which for this iridate is 343 K (**Fig.4b**). Despite the nonmetallic ground state, $\gamma$ of this iridate is estimated to be 20 mJ/mole K$^2$ at $\mu_oH=0$ T, suggesting that the finite density of states near the Fermi level arises from localization due to disorder. Interestingly, $\gamma$ is significantly greater than the 3 mJ/mole K$^2$ for SrIrO$_3$ **[17]**.

Furthermore, C(T) at low temperatures changes sensitively at relatively low magnetic field H, particularly below 1 K, as illustrated in **Fig. 4b** where C/T varies drastically with H. The approximate fitting of the data in **Fig.4b** to C(T)/T = $\gamma$ + $\beta T^2$ for 1 K <T < 4 K generates $\gamma$ as a function of H as shown in **Fig. 4c**. $\gamma$ rises pronouncedly from 20 mJ/mole K$^2$ at $\mu_oH=0$ T to 41 mJ/mole K$^2$ at $\mu_oH=2$ T before decreasing with increasing H. The origin of the unusual field-dependence of $\gamma$ is unclear but it could be due to in-gap states and/or a field-induced magnetic order that shifts up to higher temperature with increasing H (**Fig.4b**).

The octahedral crystal field splits the bands into $e_g$ and $t_{2g}$-like orbitals; there are twelve $t_{2g}$ and eight $e_g$-like bands because of the distortion of the IrO$_6$ octahedra. Our first-principles calculations for both Sr$_{0.94}$Ir$_{0.78}$O$_{2.68}$ and stoichiometric SrIrO$_3$ using the LSDA+SOI+U method with U=3eV result in different band topologies (**Figs. 5a** and **5b**). The band structure for SrIrO$_3$ is in agreement with that reported in **Ref. 10**. The calculations illustrate an insulating ground state in Sr$_{0.94}$Ir$_{0.78}$O$_{2.68}$ that emerges from a metallic state in SrIrO$_3$ as the Ir and O deficiencies break some pathways of electron hopping along the c-axis in Sr$_{0.94}$Ir$_{0.78}$O$_{2.68}$. A narrow charge gap of 0.081 eV for Sr$_{0.94}$Ir$_{0.78}$O$_{2.68}$ is predicted, which is somewhat larger but qualitatively consistent with 0.027 eV estimated from $\rho_c$ below T$_N$ or for 125 < T < 185 K. According to the calculations, the Ir- and O-deficiencies alter the $IrO_6$ octahedra, leading to three distinct Ir sites denoted by A, B and C, namely, A: $IrO_6$ octahedra remain intact, B: $IrO_6$ octahedra remain but with a broken Ir-O-Ir bond along the c- or z-direction, and C: $IrO_6$ octahedra evolve to $IrO_5$ prisms because of a missing O (**Fig.5c**). For the site A, $J_{eff}=3/2$ and $J_{eff}=1/2$ states with a magnetic moment of 0.19 $\mu_B$/Ir are anticipated; for the site B, the broken Ir-O-Ir bond along the z-direction renders a fully occupied $d_{xy}$ orbital (the lowest state) and half filled $d_{yz}$ and $d_{xz}$ orbitals. This may explain the unexpected anisotropic resistivity with $\rho_c$ showing the insulating behavior (**Fig. 3b**). In the site C the $IrO_5$ prisms lead to fully filled $d_{yz}$ and $d_{xz}$ orbitals, carrying no magnetic moment. Electron hopping between the sites A and B is anticipated, making an AFM state the energetically most favorable state (see **Fig.5d**). It is worthy mentioning that the rotation of $IrO_6$ octahedra within the basal plane is a structural signature of $Sr_2IrO_4$ and $Sr_3Ir_2O_7$ **[26-30]** and is critical to the AFM ground state **[40]**. The AFM state in $Sr_{0.94}Ir_{0.78}O_{2.68}$ might be in part a result of the increased in-plane rotation of $IrO_6$ octahedra, 9.54° (compared to 8.75° for $SrIrO_3$) whereas the decreased out-of-plane tilt, which is absent in n=1 and 2, may not be as critical in determining the ground state.

This work reveals the simultaneous AFM and MI transitions that illustrate a direct correlation between the AFM transition and charge gap in the iridate, which is conspicuously absent in $Sr_2IrO_4$. The contrasting ground states in isostructural $Sr_{0.94}Ir_{0.78}O_{2.68}$ and $SrIrO_3$ highlight the critical role of the lattice degrees of freedom that along with the delicate interplay of the SOI and U needs to be adequately addressed both experimentally and theoretically. In particular, the elusive superconductivity in the iridates, despite the apparent similarities between the cuprates and iridates, might be in part due to the ultra-high sensitivity to lattice distortions in the SOI-coupled iridates; the lattice-dependence in the cuprates is much weaker since the SOI, which anchors physical properties to the lattice in general, is negligible.

**Acknowledgements** GC is very thankful for enlightening conversations with Profs. Hae-Young Kee and Yong-Baek Kim. This work was supported by NSF through grant DMR-1265162 0856234 and the Department of Energy (BES) through grant No. DE-FG02-98ER45707 (PS). G.W. acknowledges support by Natural Science Foundation of China via Grant No.11525417

* Email: cao@uky.edu

**References**

1. B. J. Kim, Hosub Jin, S. J. Moon, J.-Y. Kim, B.-G. Park, C. S. Leem, Jaejun Yu, T. W. Noh, C. Kim, S.-J. Oh, V. Durairai, G. Cao, and J.-H. Park, Phys. Rev. Lett. **101**, 076402 (2008)

2. S.J. Moon, H. Jin, K.W. Kim, W.S. Choi, Y.S. Lee, J. Yu, G. Cao, A. Sumi, H. Funakubo, C. Bernhard, and T.W. Noh, Phys. Rev. Lett. **101**, 226402 (2008)

3. B.J. Kim, H. Ohsumi, T.Komesu, S.Sakai, T. Morita, H. Takagi, T. Arima, Science **323**, 1329 (2009)

4. *"Frontiers of 4d- and 5d- Transition Metal Oxides"*, Gang Cao and Lance E. De Long, *World Scientific*, Singapore, 2013

5. William Witczak-Krempa, Gang Chen, Yong Baek Kim, Leon Balents, Annual Review of Condensed Matter Physics, Vol. 5: 57-82 (2014)

6. Q. Wang, Y. Cao, J. A. Waugh, T. F. Qi, O. B. Korneta, G. Cao, and D. S. Dessau, *Phys. Rev B* **87** 245109 (2013)

7. Jeffrey G. Rau, Eric Kin-Ho Lee, and Hae-Young Kee, Annu. Rev. Condens. Matter Phys 7, 195 (2016)

8. D. Pesin and L. Balents, Nat. Phys. 6, 376 (2010)

9. H. Watanabe, T. Shirakawa and S. Yunoki, Phys. Rev. Lett. 105, 216410 (2010)

10. M. Ahsan Zeb and Hae-Young Kee, Phys. Rev. B 86, 085149 (2012)

11. Yige Chen Yuan-Ming Lu Hae-Young Kee, Nature Comm. doi:10.1038/ncomms7593 (2015)

12. J. M. Longo, J. A. Kafalas and R. J. Arnott, J. Solid State Chem., **3**, 174 (1971)

13. J. G. Zhao, L. X. Yang, Y. Yu, F. Y. Li, R. C. Yu, Z. Fang, L. C. Chen, and C. Q. Jin, J. Appl. Phys., **103**, 103706 2008

14. M. Bremholm, C. K. Yim, D. Hirai, E. Climent-Pascual, Q. Xu, H. W. Zandbergen, M. N. Ali and R. J. Cava, J. Mater. Chem., **22**, 16431 (2012)

15. I. Qasim, B. J. Kennedy and M. Avdeev, J. Mater. Chem. A, **1**, 3127 (2013)

16. I. Qasim, B. J. Kennedy, and M. Avdeev, J. Mater. Chem. A, **1**, 13357 (2013)

17. P. E. R. Blanchard, E. Reynolds, and B. J. Kennedy, Phys. Rev. B **89**, 214106 (2014)

18. John H. Gruenewald, John Nichols, Jasminka Terzic, Gang Cao, Joseph W. Brill, and Sung S. Ambrose Seo, J. Mater. Res. **29**, 2495 (2014)

19. Y. F. Nie, P. D. C. King, C. H. Kim, M. Uchida, H. I. Wei, B. D. Faeth, J. P. Ruf, J. P. C. Ruff, L. Xie, X. Pan, C. J. Fennie, D. G. Schlom, and K. M. Shen, Phys. Rev. Lett. **114**, 016401 (2015)

20. J. Matsuno, K. Ihara, S. Yamamura, H. Wadati, K. Ishii, V. V. Shankar, Hae-Young Kee, and H. Takagi, Phys. Rev. Lett. **114**, 247209 (2015)

21. J. Liu, J. H. Chu, C. R. Serrao, D. Yi, J. Koralek, C. Nelson, C. Frontera, D. Kriegner, L. Horak, E. Arenholz et al., arXiv:1305.1732, 2013

22. D. Hsieh, F. Mahmood, D. Torchinsky, G. Cao and N. Gedik, *Phys. Rev. B* **86**, 035128 (2012)

23. Q. Li, G. Cao, S. Okamoto, J. Yi, W. Lin, B.C. Sales, J. Yan, R. Arita, J. Kuneš, A.V. Kozhevnikov, A.G. Eguiluz, M. Imada, Z. Gai, M. Pan, D.G. Mandrus, *Scientific Reports* 3, 3073 (2013)

24. G. Cao, V. Durairaj, S. Chikara, L. E. DeLong, S. Parkin, and P. Schlottmann, Phys. Rev. B **76**, 100402(R) (2007)

25. G. M. Sheldrick, Acta Crystallogr A **64**, 112 (2008)

26. J. Rodriguez-Carvajal, Physica B. **192,** 55 (1993)

27. M.K. Crawford, M.A. Subramanian, R.L. Harlow, J.A. Fernandez-Baca, Z.R. Wang, and D.C. Johnston, Phys. Rev. B **49**, 9198 (1994)

28. R.J. Cava, B. Batlogg, K. Kiyono, H. Takagi, J.J. Krajewski, W. F. Peck, Jr., L.W. Rupp, Jr., and C.H. Chen, Phys. Rev. B **49**, 11890 (1994)

29. M.A. Subramanian, M.K. Crawford, R. L. Harlow, T. Ami, J. A. Fernandez-Baca, Z.R. Wang and D.C. Johnston, Physica C **235**, 743 (1994)

30. G. Cao, Y. Xin, C. S. Alexander, J.E. Crow and P. Schlottmann, *Phys. Rev. B* 66, 214412 (2002)

31. G. Cao, J. Bolivar, S. McCall, J.E. Crow, and R. P. Guertin, Phys. Rev. B **57**, R11039 (1998)

32. G. Cao, J.E. Crow, R.P. Guertin, P. Henning, C.C. Homes, M. Strongin, D.N. Basov, and E. Lochner, *Solid State Comm.* 113, 657 (2000)

33. G. Cao, W.H. Song, Y.P. Sun and X.N. Lin, *Solid State Comm.* **131**, 331 (2004)

34. J. H. Mooij, *Phys. Status Solidi* A **17**, 521 (1973)

35. C. A. Balseiro and L. M. Falicov, *Phys. Rev. B* **20**, 4457 (1979)

36. A. M. Gabovich, A. I. Voitenko, T. Ekino, M. S. Li, H. Szymczak, and M. Pekala, *Adv. Cond. Matter Phys.* **2010**, 681070 (2010)

37. M. Weger and N.F. Mott, *J. Phys. C: Solid St. Phys*. **18**, L201 (1985)

38. J. C. Wang, S. Aswartham, F. Ye, J. Terzic, H. Zheng, Daniel Haskel, Shalinee Chikara, Yong Choi, P. Schlottmann, S. J. Yuan[1] and G. Cao, *Phys. Rev. B* **92**, 214411 (2015)

39. J. Terzic, H. Zheng, S. J. Yuan and G. Cao, unpublished, 2015

40. G. Jackeli and G. Khaliulin, Phys. Rev. Lett. **102**, 017205 (2009)

**Captions**

**Fig. 1.** The crystal structure of perovskite $Sr_{0.94}Ir_{0.78}O_{2.68}$. Legends: green: Sr, grey: Ir, and red: Oxygen. Note there are 3.08 Ir atoms in a unit cell, as compared to 4 in $SrIrO_3$.

**Fig. 2. (a)** The temperature dependence at $\mu_oH=0.5$ T of the magnetic susceptibility $\chi$ for a-axis $\chi_a$ and c-axis $\chi_c$, and $\chi_a^{-1}$ (right scale) for $Sr_{0.94}Ir_{0.78}O_{2.68}$. **(b)** The isothermal magnetization $M_a$ (blue) and $M_c$ (red) up to 7 Tesla. **(c)** The temperature dependence of $\chi_a$ for $Sr_{n+1}Ir_nO_{3n+1}$ with n=1, 2 and $\infty$. Note $\theta_{CW}$ and $\mu_{eff}$ are the Curie-Weiss temperature and effective moment, respectively.

**Fig. 3.** The temperature dependence of **(a)** the a-axis resistivity $\rho_a$ and the c-axis resistivity $\rho_c$ (right scale) for $Sr_{0.94}Ir_{0.78}O_{2.68}$, **(b)** $\rho_a$ for $Sr_{n+1}Ir_nO_{3n+1}$ with n=1, 2 and $\infty$, and **(c)** the Hall resistivity $\rho_H$ for $Sr_{0.94}Ir_{0.78}O_{2.68}$. **Inset** in **(a)**: ln $\rho_c$ vs. $T^{-1}$. **Inset** in **(c)**: the magnetic field dependence of $\rho_H$ at a few representative temperatures.

**Fig. 4. (a)** The specific heat C(T) for 100 <T<250 K. **(b)** C(T)/T vs. $T^2$ for 0.05<T<4 K at a few representative magnetic fields. **(c)** C(T)/T or $\gamma$ vs. $\mu_oH$ for $Sr_{0.94}Ir_{0.78}O_{2.68}$.

**Fig. 5.** Comparisons of the band structures calculations using LSDA+U+SO (U=3 eV) for **(a)** $Sr_{0.94}Ir_{0.78}O_{2.68}$ and **(b)** stoichiometric $SrIrO_3$; the Fermi level is at 0. **(c)** The schematic structure for $Sr_{0.94}Ir_{0.78}O_{2.68}$ illustrates three different Ir ion sites due to the non-stoichiometry: A ($5d^5$), B ($5d^4$) and C ($5d^4$) (see text), and **(d)** the corresponding 5d-orbitals and electron hopping between the orbitals.

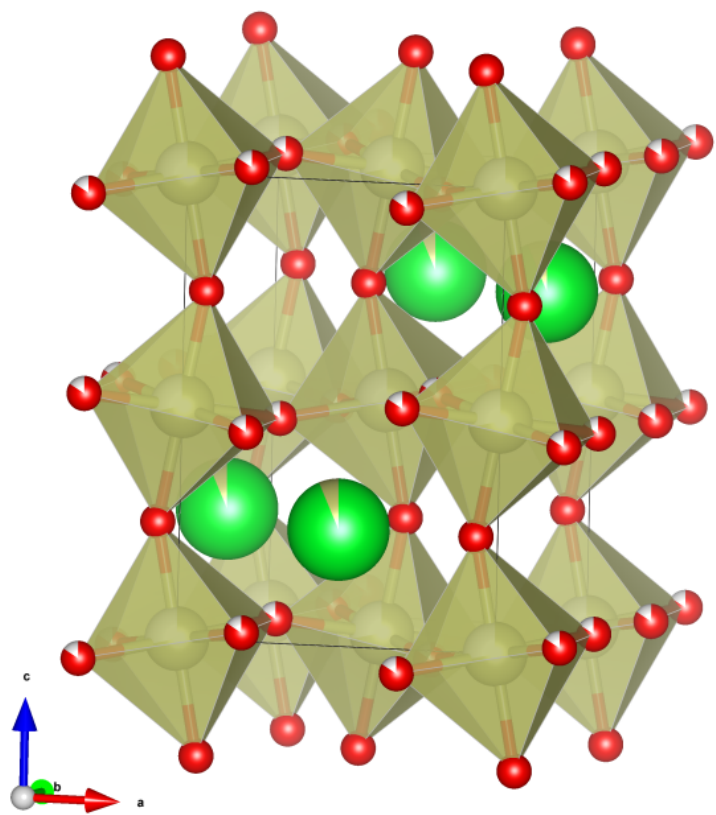

Fig.1

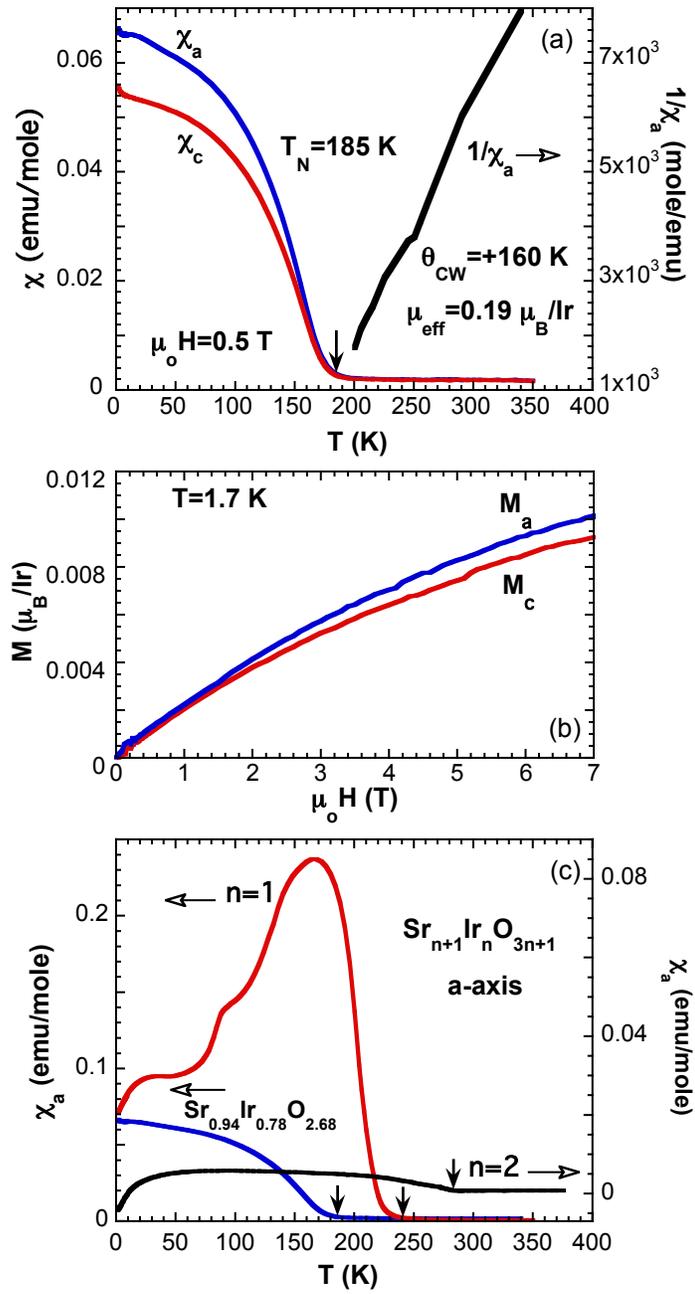

Fig. 2

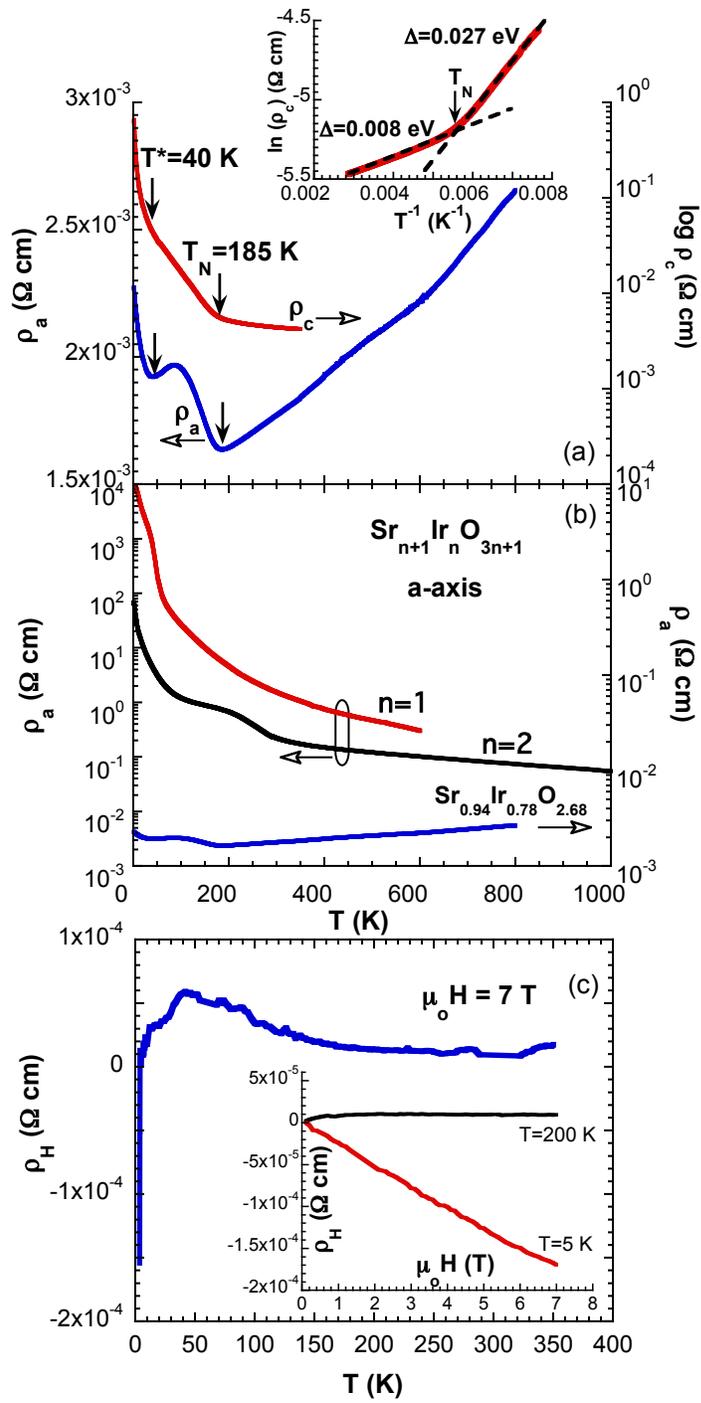

Fig. 3

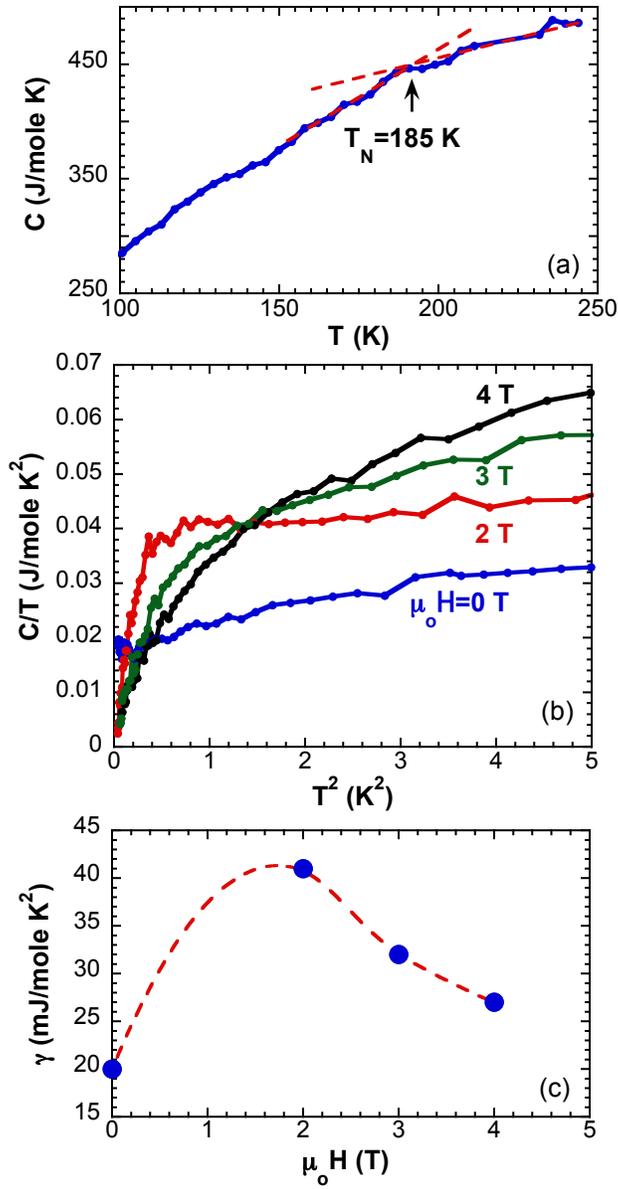

Fig. 4

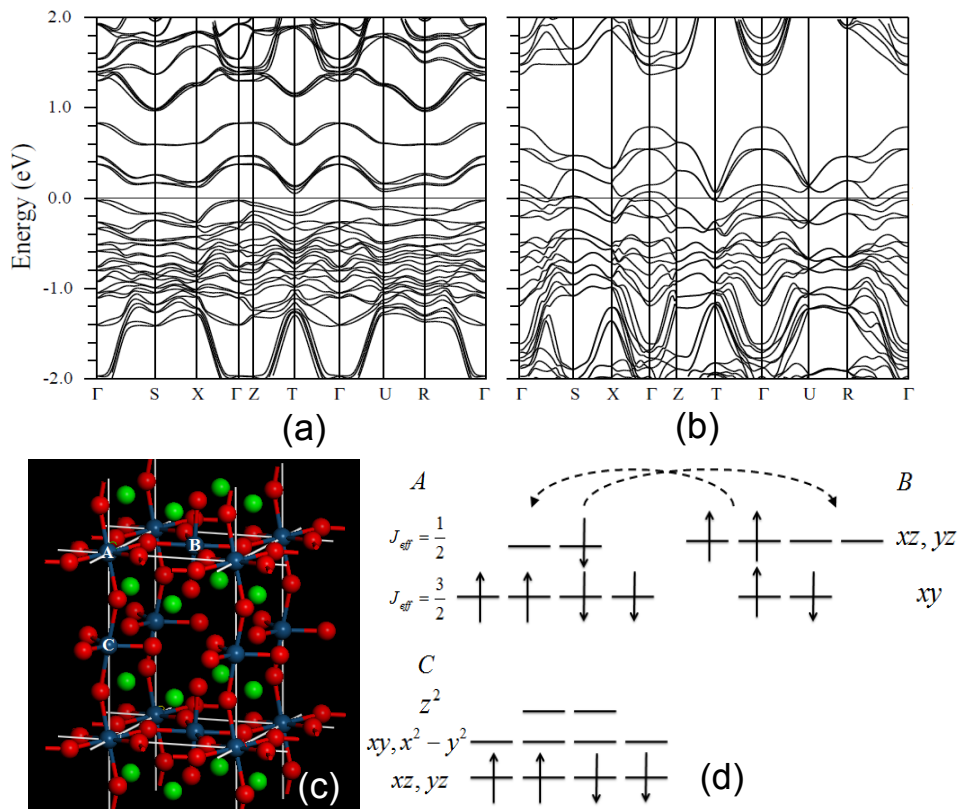

Fig.5